\documentclass[aps,prl,twocolumn,amsmath,amssymb]{revtex4}
\usepackage{graphicx}
\usepackage{amsmath}
\usepackage{color}
\usepackage[normalem]{ulem}
\begin{document}

\title{Atom-Dimer Scattering and Stability of Bose and Fermi Mixtures}
\author{Xiaoling Cui}
\affiliation{Beijing National Laboratory
for Condensed Matter Physics, Institute of Physics, Chinese Academy
of Sciences, Beijing, 100190, People's Republic of China}
\date{\today}
\begin{abstract}
Motivated by a recent experiment by ENS group on the mixture of Bose and Fermi superfluids (arxiv:1404.2548), we investigate the effective scattering between a bosonic atom and a molecule(dimer) of fermion atoms. It is found that the mean-field prediction of the atom-dimer scattering length ($a_{ad}$), as simply given by the boson-fermion scattering length ($a_{bf}$), generically fails. Instead, $a_{ad}$ crucially depends on the ratio between $a_{bf}$ and $a_{ff}$ (the fermion-fermion scattering length), and in addition log-periodically depends on the three-body parameter. We identify the universal parameters in characterizing $a_{ad}$ for a wide range of $a_{ff}$ in the molecular side of the fermion-fermion Feshbach resonance, and further demonstrate that the atom-dimer many-body system can become unstable against either phase separation or collapse as tuning $a_{ff}$. Our results have some implications to the ENS experiment.
\end{abstract}
\maketitle

The dilute ultracold atomic gases with highly tunable interactions provide an ideal platform for studying the fundamental yet challenging few-body problems, among which the Efimov physics undoubtedly takes a prominent place due to its intriguing properties\cite{Efimov, Braaten}. In 1970's, Efimov established two important laws, the scaling law and the radial law, for the bound states of three identical bosons interacting with a single s-wave scattering length $a_s$\cite{Efimov}.
The scaling law predicts the discrete scaling symmetry, i.e., a change of $a_s$ by a scaling factor $\chi=e^{\pi/s_0}$ (where $s_0=1.00624$) corresponds to the energy changed by a factor $\chi^{-2}$.
%   predicts that the three-body bound states are developed from the scattering threshold at $a_s<0$ side with critical $a_s$ separated by a universal scaling factor $\xi=e^{\pi/s_0}=22.7$ (where $s_0=1.00624$), and merge into the atom-dimer continuum at $a_s>0$ side with $a_s$ following the same scaling $\xi$; at diverging $a_s$ there are infinitely many of such bound states and their binding energies are universally spaced by scaling factor $\xi^2$.
The radial law predicts that at $a_s>0$ side, the atom-dimer scattering length $a_{ad}$ can be parametrized by $a_s$ and a three-body parameter measuring the short-range interactions among the three particles, and $a_{ad}$ diverges periodically as varying $a_s$ by a scaling factor $\chi$.
These predictions have been successfully verified in cold atom experiments on various homo- and hetero-nuclear systems, by observing the enhanced three-body recombination at $a_s<0$\cite{Efimov_Exp0,Efimov_Exp1,Efimov_Exp1bu,Efimov_Exp2,Efimov_Exp3,Efimov_Exp4,Efimov_Exp5,Efimov_Exp9,Efimov_Exp10}, the atom-dimer loss resonance
%and the interference minima of three-body recombination\cite{..}
at $a_s>0$\cite{Efimov_Exp6,Efimov_Exp7,Efimov_Exp8,Efimov_Exp11} , the Efimov spectrum measured from radio-frequency spectroscopy\cite{rf_1,rf_2}, and recently the successive three-body loss peaks directly confirming the discrete scaling symmetry\cite{scaling_1,scaling_2,scaling_3}.

Recently, the ENS group has reported a new breakthrough in realizing a mixture of Bose-Einstein condensation and fermionic superfluidity using Lithium isotopes\cite{Salomon}. In this experiment, the fermion-fermion scattering length ($a_{ff}$) can be tuned over several order of magnitude via a Feshbach resonance, while the boson-fermion scattering length ($a_{bf}$) almost stays static. This brings new challenges to the few-body physics as mainly in following two aspects. First, in the presence of more than one scattering lengths, the original Efimov predictions for single $a_s$ could be greatly affected. Secondly, given the realized extremely low temperature, the few-body physics will no doubt fundamentally influence the low-energy collective phenomena in a dilute many-body system, which have been rarely discussed before in this setup. Also considering a variety of multi-component systems\cite{Efimov_Exp1,Efimov_Exp1bu,Efimov_Exp2,Efimov_Exp7,Efimov_Exp8,Efimov_Exp11} that can be potentially cooled down to quantum degenerate regime, it is thus imperative to investigate the few-body physics in these systems and their general consequences in a many-body environment.

With above motivations, in this work we study the effective scattering between a bosonic atom and a dimer of two fermions, with tunable $a_{ff}$ and non-varying $a_{bf}$.
%($>0$ or $<0$) as well as the many-body consequences in this atom-dimer system.
By establishing a generalized Efimov's radial law, we show that the atom-dimer scattering length $a_{ad}$ sensitively depends on the ratio between the two scattering lengths, $x=a_{bf}/a_{ff}$, and can even go across resonance as tuning $x$. In addition, it log-periodically depends on the three-body parameter $\kappa^*$, similar to that of identical bosons. Its general formula is
\begin{equation}
\frac{a_{ad}}{a_{bf}}=C_1(x)+C_2(x) \cot\left[s_0\ln(\kappa^*|a_{bf}|)+\Phi(x)\right]. \label{a_ad}
\end{equation}
Here $C_1,\ C_2, \ \Phi$ are all universal functions in terms of $x$. We numerically verify this formula and extract those universal functions for a wide range of $x$  (taking equal mass of boson and fermion for instance). Moreover, we show that the mean-field prediction of $a_{ad}$, which is proportional to $a_{bf}$ as determined from the boson-fermion density-density interaction, generically fails for typical cold atoms systems with short-range interactions. Furthermore, based on Eq.(\ref{a_ad}), %we further demonstrate that 
we show that the stability of the atom-dimer many-body system can be greatly altered by tuning $x$, where the homogenous mixture can become unstable against phase separation or collapse.
%boson-boson or fermion-fermion interactions.
We identify the according phase diagram for several typical values of $a_{bf}$ and $\kappa^*$. Finally we point out some implications of our results to the ENS experiment\cite{Salomon}.

% atom loss and collective mode measurement in  
%Our results , and discuss the significance of our results in a general background of mixture systems.

{\it Model.} Considering two distinguishable fermions at positions ${\bf r}_1,\ {\bf r}_2$ with mass $m_f$ and one boson at ${\bf r}_3$ with mass $m_b$, the Hamiltonian can be written as $\hat{H}=\hat{H_0}+\hat{U}$,
\begin{equation}
\hat{H_0}=-\frac{\nabla_{\bf r}^2}{2\mu}-\frac{\nabla_{\boldsymbol\rho}^2}{2\mu_{\rho}};\ \
\hat{U}=U_{ff}\delta({\bf r})+U_{bf}\big[ \delta({\boldsymbol\rho}+\frac{{\bf r}}{2})+\delta({\boldsymbol\rho}-\frac{{\bf r}}{2})\big].
\end{equation}
Here ${\bf r}={\bf r}_2-{\bf r}_1$ and $\boldsymbol\rho={\bf r}_3-({\bf r}_1+{\bf r}_2)/2$ respectively describe the relative motion between two fermions and between the boson and the center of mass of fermions; the corresponding masses are $\mu=m_f/2$ and $\mu_{\rho}=2m_fm_b/(2m_f+m_b)$. $U_{bf}$ ($U_{ff}$) is the bare interaction between boson-fermion (fermion-fermion) and can be related to $a_{bf}$ ($a_{ff}$) via $\frac{1}{U_{bf}}=\frac{\bar{\mu}}{2\pi a_{bf}}-\frac{1}{V}\sum_{\bf k}\frac{2\bar{\mu}}{k^2}$ $ \Big(\frac{1}{U_{ff}}=\frac{\mu}{2\pi a_{ff}}-\frac{1}{V}\sum_{\bf k}\frac{2\mu}{k^2} \Big)$, where $\bar{\mu}=m_bm_f/(m_b+m_f)$, and $V$ is the volume.

\begin{figure}[h]
\includegraphics[height=5cm]{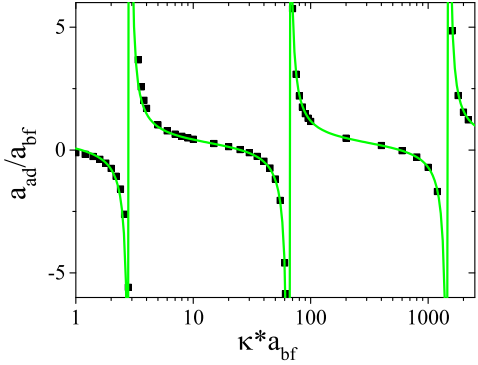}
\caption{(Color online) Verification of universal formula for $a_{ad}$ (in the unit of $a_{bf}$, see Eq.(\ref{a_ad})) at $x=a_{bf}/a_{ff}=2$. The square points show numerical results by solving Eqs.(\ref{eq1},\ref{eq2}) for different momentum cutoffs $\kappa^*$. The green curve is the fit to Eq.(\ref{a_ad}) with $s_0=1.00624,\ C_1=0.29,\ C_2=0.41,\ \Phi=-0.34\pi$. %The first three resonances of $a_{ad}$ occur at $\kappa^* a_{bf}=$.
The fit gets more accurate for larger $\kappa^* a_{bf}$. } \label{fig1}
\end{figure}

Now we study the atom-dimer elastic scattering between the boson and the dimer of fermions. The associated energy is $E=-\epsilon_{ff}$, where $\epsilon_{ff}=1/(2\mu a_{ff}^2)$ is the dimer binding energy. We solve the wave function $\Psi$ using the Lippman-Schwinger equation $|\Psi\rangle=\hat{G}_0\hat{U}|\Psi\rangle$, with $\hat{G}_0=1/(E-\hat{H}_0+i\epsilon)$ the non-interacting Green's function. Introducing three auxiliary functions in
\begin{equation}
\langle {\bf r}, \boldsymbol\rho | \hat{U} |\Psi\rangle= f(\boldsymbol\rho)\delta({\bf r})+g_{+}({
\bf r}) \delta({\boldsymbol\rho}+\frac{{\bf r}}{2})+ g_{-}({
\bf -r})\delta({\boldsymbol\rho}-\frac{{\bf r}}{2}), \label{U}
\end{equation}
%and according to the L-S equation
we arrive at three coupled equations in terms of $f,\ g_{+}$ and $g_{-}$, which correspond to implementing the boundary conditions to $\Psi$ respectively at ${\bf r}\rightarrow 0,\ \boldsymbol\rho+\frac{{\bf r}}{2}\rightarrow 0$ and $\boldsymbol\rho-\frac{{\bf r}}{2}\rightarrow 0$\cite{supple}. For instance, we have
\begin{equation}
\lim_{{\bf r}\rightarrow 0}\Psi({\bf r}, \boldsymbol\rho)\sim f(\boldsymbol\rho) (1/r-1/a_{ff}). %(\frac{1}{r}-\frac{1}{a_{ff}}).
\end{equation}
The physical meaning of $f-$function is thus transparent, that it effectively describes the relative motion between the atom (boson) and the dimer (fermions) and can be referred to as atom-dimer scattering wave function. The atom-dimer scattering length, $a_{ad}$, can be extracted from the Fourier transformation of $f-$function in ${\bf k}$ space,
\begin{equation}
f({\bf k})=(2\pi)^3\delta({\bf k})-\frac{4\pi a_{ad}({\bf k})}{k^2},
\end{equation}
with $a_{ad}\equiv a_{ad}(0)$ given by on-shell atom-dimer T-matrix element at zero momentum.

The atom-dimer scattering with unequal scattering lengths has been studied before in three-component Li6 system\cite{Braaten2}. For our system, where the boson-fermion interactions are described by a single $a_{bf}$, we have $g_{+}=g_{-}\equiv g$ and finally  we obtain two coupled integral equations for $a_{ad}({\bf k})$ and $g({\bf k})$:
%\begin{widetext}
%\begin{eqnarray}
%&&-\frac{4\pi a_{ad}({\bf k})}{k^2} \left( \frac{1}{U_{ff}} - \frac{1}{V} \sum_{\bf q} \frac{1}{E- \frac{q^2}{2\mu}-\frac{k^2}{2\mu_{\rho}}+i\epsilon} \right)\
%=  \frac{1}{V} \sum_{\bf p} \frac{2g({\bf p})}{E- \frac{({\bf k}/2+{\bf p})^2}{2\mu}-\frac{k^2}{2\mu_{\rho}}+i\epsilon}; \label{eq1} \\
%&&g({\bf p})  \left( \frac{1}{U_{bf}} - \frac{1}{V} \sum_{\bf k} \frac{1}{E- \frac{({\bf k}/2+{\bf p})^2}{2\mu}-\frac{k^2}{2\mu_{\rho}}+i\epsilon} \right)
%= \frac{1}{E-\frac{p^2}{2\mu}} + \frac{1}{V} \sum_{\bf k} \left( \frac{-4\pi a_{ad}({\bf k})}{k^2 (E- \frac{({\bf k}/2+{\bf p})^2}{2\mu}-\frac{k^2}{2\mu_{\rho}}+i\epsilon) }+  \frac{g({\bf k})}{ E- \frac{({\bf p}-{\bf k})^2}{8\mu}-\frac{({\bf p}+{\bf k})^2}{2\mu_{\rho}}+i\epsilon } \right). \label{eq2}
%\end{eqnarray}
%\end{widetext}
%
\begin{eqnarray}
&&-\frac{4\pi a_{ad}({\bf k})}{k^2} \left( \frac{1}{U_{ff}} - \frac{1}{V} \sum_{\bf q} \frac{1}{E- \frac{q^2}{2\mu}-\frac{k^2}{2\mu_{\rho}}+i\epsilon} \right)\nonumber\\
&=&  \frac{1}{V} \sum_{\bf p} \frac{2g({\bf p})}{E- \frac{({\bf k}/2+{\bf p})^2}{2\mu}-\frac{k^2}{2\mu_{\rho}}+i\epsilon}; \label{eq1} \\
&&g({\bf p})  \left( \frac{1}{U_{bf}} - \frac{1}{V} \sum_{\bf k} \frac{1}{E- \frac{({\bf k}/2+{\bf p})^2}{2\mu}-\frac{k^2}{2\mu_{\rho}}+i\epsilon} \right)
\nonumber\\
&=&\frac{1}{E-\frac{p^2}{2\mu}} + \frac{1}{V} \sum_{\bf k} \left( \frac{-4\pi a_{ad}({\bf k})}{k^2 (E- \frac{({\bf k}/2+{\bf p})^2}{2\mu}-\frac{k^2}{2\mu_{\rho}}+i\epsilon) }+ \right. \nonumber\\
&& \ \ \ \ \ \ \ \ \ \ \ \  \left. \frac{g({\bf k})}{ E- \frac{({\bf p}-{\bf k})^2}{8\mu}-\frac{({\bf p}+{\bf k})^2}{2\mu_{\rho}}+i\epsilon } \right). \label{eq2}
\end{eqnarray}

Eqs.(\ref{eq1},\ref{eq2}) generally apply to arbitrary mass ratios $\eta=m_f/m_b$ and arbitrary $x=a_{bf}/a_{ff}$. In this work we consider $m_b=m_f=m \ (\eta=1)$ for instance, and mainly focus on the region $x\in(1,+\infty)\bigcup(-\infty,0)$, where the fermion-fermion dimer is the ground state dimer in the three-body system\cite{footnote_loss}. % and accordingly $f-$ and $g-$functions are all real.

%{\color{red} Furthermore, we note that the atom-dimer scattering with unequal scattering lengths was previously considered in three-component Li6 system\cite{Braaten}, where the result corresponds to a special case in our work (see Eq.(8) later). }

{\it Generalized Efimov's radial law.} Before proceeding with numerical solutions from Eqs.(\ref{eq1},\ref{eq2}), we first prove that $a_{ad}$ can be parameterized explicitly by a few parameters and follows the universal form as Eq.(\ref{a_ad}). This is a straightforward generalization of Efimov's radial law\cite{Efimov} to the case of multiple scattering lengths. For the proof, we only illustrate the main idea here, which contains two essential ingredients: (here $R\sim\sqrt{\sum_{i<j}|{\bf r}_i-{\bf r}_j|^2}$ is the hyperradius; $r_0$ is the range of interaction potential)
%; $(H,\xi)$ is the polar variables in ())

(A) In the scale-invariant regime $r_0\ll R\ll |a_{bf}|,\ |a_{ff}|$, the three-body potential is identical to that with three divergent scattering lengths and follows an attractive $1/R^2$ form. The resulted wave function is formulated as $\sin(s_0\ln R+\theta)$, where $\theta$ is determined by the boundary condition at short-range $R\sim r_0$ and thus incorporates the three-body parameter $\kappa^*$. This formulation is unchanged in the presence of multiple scattering lengths. %, as long as they are all in the universal regime, i.e., $|a_{bf}|,\ |a_{ff}|\gg r_0$

(B) The atom-dimer elastic scattering is given by the asymptotic behavior of above three-body wave function at $R\gg |a_{bf}|,\ |a_{ff}|$. In order to identify that, one has to consider the intermediate regime of $R\sim|a_{bf}|,\ |a_{ff}|$. %, where the pair-wise interactions take effect.
The interaction in this regime can be effectively considered as certain potential barrier and thus is characterized by reflection and transmission coefficients, which are dimensionless numbers and can be parametrized by scattering lengths and the wave-vector of the wave function. Given the barrier characteristics, one can formulate explicitly the evolution of wave function from $ R\ll |a_{bf}|,\ |a_{ff}|$ to $R\gg |a_{bf}|,\ |a_{ff}|$, and extract the atom-dimer phase shift or $a_{ad}$ in the latter regime. The general form is as Eq.(\ref{a_ad}). Compared to the original Efimov's radial law\cite{Efimov}, here the specialty is that we need multiple scattering lengths, $a_{bf}$ and $a_{ff}$, to characterize the barrier, and this leads to the dependence of those characterization coefficients ($C_1,\ C_2,\ \Phi$) on the scattering length ratio $x$.

The universal formula of $a_{ad}$ as Eq.(\ref{a_ad}) can be numerically verified by exactly solving Eqs.(\ref{eq1},\ref{eq2}) for different momentum cutoffs $\kappa^*$. In Fig.1, we show $a_{ad}/a_{bf}$ as functions of $\kappa^*a_{bf}$ for a typical value of $x(=2)$, and the log-periodic fit to Eq.(\ref{a_ad}) is excellent (especially at large $\kappa^* a_{bf}$). We have also computed for other positive and negative $x$ and equally verified Eq.(\ref{a_ad}).

%{\it Breakdown of mean-field theory.}

The validity of Eq.(1) in turn implies the breakdown of mean-field theory, which gives $a_{ad}=8/3 a_{bf}$ by equating the density-density interaction between bosons and fermions with that between bosons and dimers\cite{supple}. %of boson-fermion with that of boson-dimer.  
%determined from the density-density interaction between bosons and fermions. 
In fact, this result corresponds to approximating $g({\bf p})$ as $\frac{2\pi a_{bf}/\bar{\mu}}{E-p^2/(2\mu)}$ in Eq.(\ref{eq2}) and also enforcing $\kappa^*\gg 1/a_{ff}$ in Eq.(\ref{eq1})\cite{supple}. Overall it requires $|a_{bf}|\ll 1/\kappa^* \ll a_{ff}$, where $1/\kappa^*\sim r_0$ is typically the range of interacting potential. Apparently under this requirement the system cannot be in the scale-invariant regime (which violates (A)) and one must take into account the finite-range effect. Given typical cold atoms systems with $r_0$ much shorter than the scattering lengths, we conclude that the mean-field prediction to $a_{ad}$ generically fails.

\begin{figure}[h]
\includegraphics[width=8.5cm, height=6.5cm] {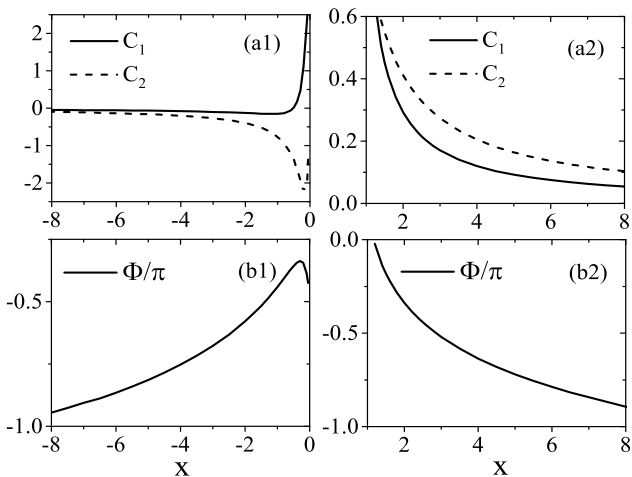}
\caption{Universal parameters $C_1,\ C_2$ (upper panel) and $\Phi$(lower panel) as functions of $x$.
%$x\in(1,+\infty)\bigcup(-\infty,0)$.
In (a1,b1), $x<0$; in (a2,b2), $x>1$.   } \label{fig2}
\end{figure}

{\it Universal functions.} After establishing Eq.(\ref{a_ad}), one can numerically extract the three universal parameters $C_1,\ C_2,\ \Phi$ as functions of $x$. The results for both negative and positive $x$ are shown in Fig.2. Several conclusions can be drawn as follows.

First, Fig.2(a1,a2) show that both the amplitudes of $C_1,\ C_2$ decay at large $|x|$, which means that $|a_{ad}/a_{bf}|$ is generally very small (except near resonance) in the limit of $|a_{bf}|\gg a_{ff}\ (|x|\gg 1)$. This actually indicates a unitary regime when $a_{bf}\rightarrow \pm \infty$, where $a_{ad}$ approaches a universal value that only depends on $a_{ff}$ and $\kappa^*$ but not $a_{bf}$ any more. The existence of this regime is verified in Fig.3(a), by plotting $a_{ad}$ for fixed $\kappa^*$ and $a_{ff}$ while changing $a_{bf}$ to $-\infty$ or $+\infty$. Exactly at $|a_{bf}|= \infty$, $a_{ad}$ can be formulated as
\begin{equation}
\frac{a_{ad}}{a_{ff}}=c_1+c_2 \cot\left[s_0\ln(\kappa^*a_{ff})+\phi\right], \ \ \ \ (|a_{bf}|=\infty) \label{a_ad_2}
\end{equation}
In Fig.3(b), we show the log-periodic dependence of $a_{ad}/a_{ff}$ on $\kappa^*$, and extract the three universal parameters as $(c_1,\ c_2,\ \phi)=(-0.15,\ -0.31,\ -0.25\pi)$. Similar formula as Eq.(\ref{a_ad_2}) was obtained previously in Li6 system with three scattering lengths\cite{Braaten2}.

\begin{figure}
\includegraphics[width=9cm] {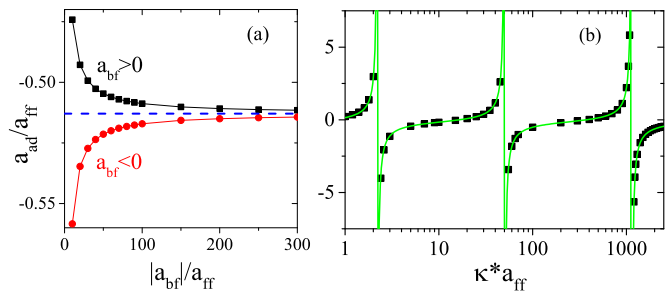}
\caption{(Color online) $a_{ad}$ in the unitary limit of $a_{bf}$. The length unit here is $a_{ff}$. (a) $a_{ad}$ as changing $a_{bf}$ to $+\infty$(black square) or $-\infty$(red circle) for fixed $\kappa^*a_{ff}=100$. The dashed horizontal line denotes the universal value $a_{ad}/a_{ff}=-0.513$ at unitarity ($|a_{bf}|=\infty$). (b) $a_{ad}$ as a function of $\kappa^*$ at $|a_{bf}|=\infty$. The green curve is the fit to Eq.(\ref{a_ad_2}). % with $s_0=1.00624,\ C_1^{\infty}=,\ C_2^{\infty}=,\ \Phi^{\infty}/\pi=$
} \label{fig3}
\end{figure}

Secondly, the dependence of the phase $\Phi$ on $x$, as shown in Fig.2(b1,b2), implies that $a_{ad}$ can also be tuned to resonance by changing $x$ while fixing $\kappa^*$. Near the resonance position $x_{res}$, we have
\begin{equation}
\frac{a_{ad}}{a_{bf}}=\frac{W}{x-x_{res}}+{\rm const},
\end{equation}
where $W=C_2(x_{res})/\Phi'(x_{res})$ is the resonance width. In Fig.4(a), we show a typical resonance structure of $a_{ad}$ as tuning $x$ with a fixed $\kappa^*$. Accordingly, the width $W$ is plotted in Fig.4(b) for both positive and negative resonance position $x=x_{res}$. It is readily seen that the absolute value of $W$ is of the order of 1 or even larger for most values of $x$, thus these resonances should be easy to resolve in cold atoms experiments given not so small $a_{bf}$. Such resonance structure of $a_{ad}$ would significantly influence the stability of atom-dimer many-body system, as demonstrated below.

\begin{figure}[h]
\includegraphics[width=9cm] {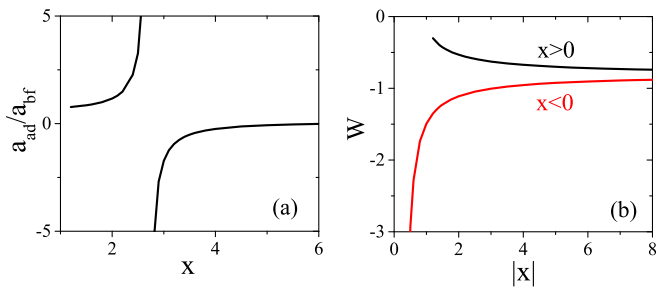}
\caption{(Color online) Resonance of $a_{ad}$ by tuning $x$. (a) $a_{ad}/a_{bf}$ as functions of $x$ for fixed $\kappa^* a_{bf}=100$. The resonance position is $x= 2.7$. (b) Resonance width $W$ for positive (black line) and negative (red line) resonance position $x$. } \label{fig4}
\end{figure}

{\it Many-body phase diagram.} For a many-body system %consisting of %(bosonic) atoms and  (fermion-fermion) dimers, 
composed by the boson condensate and the dimer(fermion-fermion) superfluid, the total energy density, ${\cal E}=E/V$, can be written as\cite{Eint}
\begin{equation}
{\cal E}=\frac{n_b^2}{2}\frac{4\pi a_{bb}}{m}+\frac{n_d^2}{2}\frac{2\pi
a_{dd}}{m}+n_bn_d\frac{3\pi a_{ad}}{m},  \label{energy}
\end{equation}
where $n_b,\ n_d$ are respectively the density of bosons and dimers, and $a_{bb},\ a_{dd}(=0.6a_{ff}$\cite{Petrov}) are the boson-boson and dimer-dimer scattering lengths.
% and $a_{ad}$ is the boson-dimer scattering length following Eq.(\ref{a_ad}).
%second derivative of E with respect to density
The stability of a homogeneous mixture system can be examined through the second-order variation of the energy functional ${\cal E}$ with respect to the density fluctuations of the atoms and dimers\cite{book}. Following the standard analysis, we arrive at three phases in different parameter regimes: (i)homogenous mixture(MIX) at $|a_{ad}|<\xi$; (ii) phase separation(PS) at $a_{ad}>\xi$; and (iii)collapse (CL) at $a_{ad}<-\xi$, where $\xi=0.73\sqrt{a_{bb}a_{ff}}$. The system is unstable in (ii) and (iii) towards a spatial separation of atoms and dimers and the formation of a denser state containing both components\cite{footnote1}. The phase diagram can be deduced accordingly in $(a_{bb},a_{ff})$ plane given that $a_{ad}$ is known from few-body solutions (Eq.(\ref{a_ad})).

In Fig.5, we present the phase diagrams for both positive and negative $a_{bf}$ with a fixed $\kappa^*$. Due to the resonance structure of $a_{ad}$ as tuning $x$(as shown in Fig.4(a)), the phase diagram is very rich.
% among the three phases illustrated above.
Typically, at a given $a_{bb}$ and by increasing $|x|$, the system can go through four phases in order: MIX-PS-CL-MIX\cite{footnote}. This is in distinct contrast to the phase diagram based on the mean-field prediction $a_{ad}=8/3 a_{bf}$, where only one phase boundary exists between MIX-PS or MIX-CL (see red dashed lines). Note that the reentrance of MIX phase at large $|x|$ in Fig.5 is due to the universal behavior of $a_{ad}$ when $a_{bf}\rightarrow\infty$, which scales as $a_{ff}$ rather than keeps growing with $a_{bf}$ as mean-field predicted. In this limit, the phase diagram can be straightforwardly obtained in $(a_{bb},a_{ff})$ plane based on Eq.(\ref{a_ad_2}), which will not be shown here.

\begin{figure}[t]
\includegraphics[height=8cm] {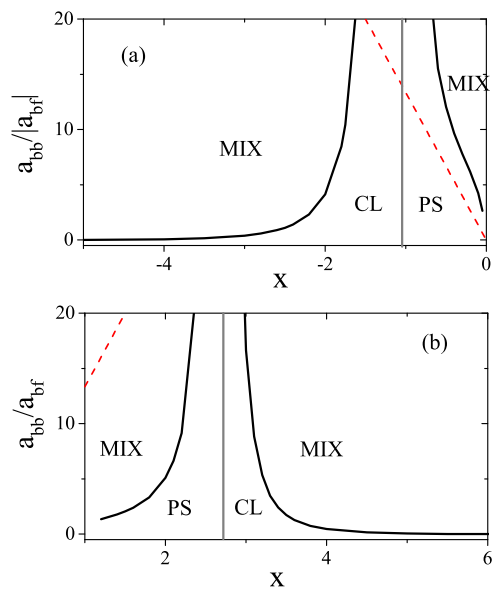}
\caption{(Color online) Many-body phase diagram of atom-dimer system 
%in the ($a_{bb}/|a_{bf}|,\ a_{bf}/a_{ff}$) plane, 
for negative(a) and positive(b) $a_{bf}$ and with a fixed $\kappa^*|a_{bf}|=100$. Three
phases are shown: homogenous mixture (MIX), phase separation (PS),
and collapse phase (CL). Black solid lines denote the phase boundaries between MIX and PS/CL. Gray vertical lines mark the atom-dimer resonances at $x=-1.21(a),\ 2.7(b)$, which also serve as PS-CL phase boundaries. Red dashed lines shows 
phase boundaries based on the mean-field prediction of $a_{ad}$; above the boundary is
MIX phase and below is CL (a) or PS (b) phase. }
\label{fig5}
\end{figure}

{\it Implications to ENS experiment\cite{Salomon}.} Our results have some implications to the ENS experiment.
%, which was mostly carried out in the $a_{bf}<|a_{ff}|$ regime\cite{Salomon}. First, for $a_{bf}<a_{ff}\ (0<x<1)$, $g({\bf k})$ (Eq.(\ref{eq2})) will have a pole at finite ${\bf k}$ and both $f({\bf k})$ and $g({\bf k})$ will become complex. Physically it is because the fermion-fermion dimer is not the ground state dimer, which has a finite life time and decays to the lower boson-fermion dimer by three-body collisions. However, the decay rate is expected to be quite low given $a_{ff}\gg a_{bf}$ and the energy gap between two dimers is sufficiently large, as the actual case in ENS experiment. On the other hand, we expect the situation will be dramatically changed if $a_{ff}$ can be tuned close to $a_{bf}$ or even smaller\cite{footnote2}. 
%{\color{red} First, ENS experiment is mostly carried out in the $a_{bf}<|a_{ff}| \ (0<x<1)$ regime\cite{Salomon}. In this regime, $g({\bf k})$ (Eq.(\ref{eq2})) will have a pole at finite ${\bf k}$ and both $f({\bf k})$ and $g({\bf k})$ will become complex. We thus expect the Bose and Fermi mixtures in ENS setup will suffer from three-body loss due to the decay of fermion-fermion dimer to boson-fermion dimer through three-body collisions.  }
We have shown that the mean-field evaluation of $a_{ad}$ generically fails. Therefore at deep molecular side of fermions, the dipole mode frequency shift of the bosons, $\delta \omega_b$, has to employ the exact solution of $a_{ad}$ (our Eq.(\ref{a_ad})) rather than Eq.(10) in Ref.\cite{Salomon}. Given the resonance structure of $a_{ad}$ as tuning $a_{ff}$ (our Fig.4(a)), $\delta \omega_b$ is also expected to exhibit similar intriguing features in this regime.  %Since the experiment shows that the mean-field theory (based on density-density interaction of bosons and fermions) works well for fermions in non-interacting limit and even near unitarity(see Fig.4 in Ref.\cite{Salomon}), 
It is promising that by measuring the deviation of $\delta\omega_b$ from mean-field evaluations (based on density-density interaction of bosons and fermions), one could see clearly how the mean-field theory breaks down as fermions gradually approaching the molecule limit and the atom-dimer picture starting to develop.
%Finally, it would be interesting problem t is also interesting to investigate how the mean-field theory (based on density-density interaction of bosons and fermions) gradually breaks down as fermions approaching the strong coupling limit and atom-dimer picture starting to develop.

{\it Concluding remark.} In conclusion, we have established a general formula (Eq.(1)) for the atom-dimer elastic scattering in the Bose and Fermi mixture, %which is found to significantly influence 
and pointed out its significant effect to the stability of a many-body system. Remarkably, by tuning the ratio between the boson-fermion and fermion-fermion scattering lengths, we show that the atom-dimer scattering length $a_{ad}$ can vary dramatically and even go across resonance. This invalidates the mean-field prediction of $a_{ad}$, and suggests 
there should be more sophisticated treatment to the interaction energy 
%more sophisticated method should be employed to treat the interaction energy 
as the system approaching the strong coupling limit of fermions. Moreover, our results show the possibility of tuning the effective interaction between a particle and a composite one, simply by adjusting the internal interaction within the latter. This scheme can be generally applied to a vast class of ultracold boson and fermion mixtures\cite{Chin}, and is expected to significantly influence the many-body physics therein.

%In conclusion, we have studied the elastic atom-dimer scattering in a mixture of Bose and Fermi superfluids, and pointed out its significant effect to the stability of a many-body system. In particular, we show that the atom-dimer interaction sensitively relies on the ratio between different interaction strengths in this system (as shown by Eq.(\ref{a_ad})). This reveals the possibility of efficiently tuning the interaction between a particle and a composite one, simply by adjusting the internal interaction within the latter. This scheme can be generally applied to a vast class of ultracold boson and fermion mixtures\cite{Chin}, and is expected to significantly influence the many-body physics in these systems.

{\it Acknowledgements.} The author thanks Tin-Lun Ho for helpful discussions. This work is supported by NSFC under Grant No. 11104158, No. 11374177, and programs of Chinese Academy of Sciences.

\clearpage

\begin{widetext}

\section{Supplementary Material}

In this supplementary material, we present details for deriving the three-body equations (Eqs.(6,7) in the main text), and also discuss the validity of mean-field prediction to atom-dimer scattering length based on these equations.

\subsection{I. Deriving three-body equations}

Based on the Lippman-Schwinger equation $|\Psi\rangle=\hat{G}_0\hat{U}|\Psi\rangle$, we have
\begin{equation}
\langle {\bf r}, \boldsymbol\rho | \hat{U} |\Psi\rangle=\langle {\bf r}, \boldsymbol\rho | \hat{U} \hat{G}_0\hat{U}|\Psi\rangle.
\end{equation}
Acting $\lim_{{\bf r}\rightarrow 0}$ onto both sides of above equation, and combining with Eq.(3) in the main text, we obtain
\begin{equation}
\frac{1}{U_{ff}} f(\boldsymbol\rho) =\int d\boldsymbol\rho' \langle 0, \boldsymbol\rho | \hat{G}_0 | 0, \boldsymbol\rho' \rangle f(\boldsymbol\rho') + \int d{\bf r'} \langle 0, \boldsymbol\rho | \hat{G}_0 | {\bf r'}, -{\bf r'}/2  \rangle g_+({\bf r'}) + \int d{\bf r'} \langle 0, \boldsymbol\rho | \hat{G}_0 | {\bf r'}, {\bf r'}/2  \rangle g_-(-{\bf r'}) . \label{f}
\end{equation} 
Similarly, acting $\lim_{\boldsymbol\rho+ {\bf r}/2 \rightarrow 0}$ or $\lim_{\boldsymbol\rho- {\bf r}/2 \rightarrow 0}$ onto both sides, we obtain another two equations:
\begin{equation}
\frac{1}{U_{bf}} g_+({\bf r}) =\int d\boldsymbol\rho' \langle {\bf r}, -{\bf r}/2 | \hat{G}_0 | 0, \boldsymbol\rho' \rangle f(\boldsymbol\rho') + \int d{\bf r'} \langle {\bf r}, -{\bf r}/2 | \hat{G}_0 | {\bf r'}, -{\bf r'}/2  \rangle g_+({\bf r'}) + \int d{\bf r'} \langle {\bf r}, -{\bf r}/2 | \hat{G}_0 | {\bf r'}, {\bf r'}/2  \rangle g_-(-{\bf r'}); \label{g+}
\end{equation} 
\begin{equation}
\frac{1}{U_{bf}} g_-(-{\bf r}) =\int d\boldsymbol\rho' \langle {\bf r}, {\bf r}/2 | \hat{G}_0 | 0, \boldsymbol\rho' \rangle f(\boldsymbol\rho') + \int d{\bf r'} \langle {\bf r}, {\bf r}/2 | \hat{G}_0 | {\bf r'}, -{\bf r'}/2  \rangle g_+({\bf r'}) + \int d{\bf r'} \langle {\bf r}, {\bf r}/2 | \hat{G}_0 | {\bf r'}, {\bf r'}/2  \rangle g_-(-{\bf r'}). \label{g-}
\end{equation} 
Eqs.(\ref{f},\ref{g+},\ref{g-}) comprise a closed set of three-body equations, from which one can solve the trimer binding energy and the atom-dimer scattering length. Actually, these equations also correspond to implementing the short-range boundary conditions to $\Psi$ respectively at ${\bf r}\rightarrow 0,\ \boldsymbol\rho+\frac{{\bf r}}{2}\rightarrow 0$ and $\boldsymbol\rho-\frac{{\bf r}}{2}\rightarrow 0$. A simplest way to see this is by rewriting $\hat{U}$ in the form of pseudo-potential. For instance, $U_{ff} \delta({\bf r})=\frac{2\pi a_{ff}}{\mu} \delta({\bf r}) \frac{\partial}{\partial r}(r\cdot)$. Then according to the definition of $f(\boldsymbol\rho)$ (Eq.(3) in the main text), we can relate it to the short-range behavior of $\Psi$: (see also Eq.(4) in the main text)
\begin{equation}
\lim_{{\bf r}\rightarrow 0}\Psi({\bf r}, \boldsymbol\rho)\sim f(\boldsymbol\rho) \left(\frac{1}{r}-\frac{1}{a_{ff}}\right). %(\frac{1}{r}-\frac{1}{a_{ff}}).
\end{equation}
Similarly, we have another two short-range boundary conditions for $\Psi$:
\begin{eqnarray}
\lim_{\boldsymbol\rho+\frac{{\bf r}}{2}\rightarrow 0 }\Psi({\bf r}, \boldsymbol\rho)&\sim& g_+({\bf r}) \left(\frac{1}{|\boldsymbol\rho+\frac{{\bf r}}{2}|}-\frac{1}{a_{bf}}\right)  ; \\
\lim_{\boldsymbol\rho-\frac{{\bf r}}{2}\rightarrow 0 }\Psi({\bf r}, \boldsymbol\rho)&\sim& g_-(-{\bf r}) \left(\frac{1}{|\boldsymbol\rho-\frac{{\bf r}}{2}|}-\frac{1}{a_{bf}}\right).
\end{eqnarray}
Therefore the physical meanings of $f,\ g_+$ and $g_-$ functions are transparent, which respectively describe the relative motion between one atom and a dimer (or center-of-mass motion) of the other two atoms. Among them, the $f-$function is responsible for the relative motion between the boson and the dimer of fermions, from which one can define the atom-dimer scattering length, $a_{ad}$, via Eq.(5) in the main text.

To extract $a_{ad} $ from Eqs.(\ref{f},\ref{g+},\ref{g-}), we work with their Frourier transformations in momentum space:
\begin{eqnarray}
f({\bf k}) \left( \frac{1}{U_{ff}} - \frac{1}{V} \sum_{\bf q} \frac{1}{E- \frac{q^2}{2\mu}-\frac{k^2}{2\mu_{\rho}}+i\epsilon} \right)
&=&  \frac{1}{V} \sum_{\bf p} \left( \frac{g_+({\bf p})}{E- \frac{({\bf k}/2+{\bf p})^2}{2\mu}-\frac{k^2}{2\mu_{\rho}}+i\epsilon} + \frac{g_-({\bf p})}{E- \frac{({\bf k}/2+{\bf p})^2}{2\mu}-\frac{k^2}{2\mu_{\rho}}+i\epsilon} \right) ; \label{f_2} \\
g_+({\bf p})  \left( \frac{1}{U_{bf}} - \frac{1}{V} \sum_{\bf k} \frac{1}{E- \frac{({\bf k}/2+{\bf p})^2}{2\mu}-\frac{k^2}{2\mu_{\rho}}+i\epsilon} \right) 
&=&\frac{1}{V} \sum_{\bf k} \left( \frac{f({\bf k})}{E- \frac{({\bf k}/2+{\bf p})^2}{2\mu}-\frac{k^2}{2\mu_{\rho}}+i\epsilon }+  \frac{g_-({\bf k})}{ E- \frac{({\bf p}-{\bf k})^2}{8\mu}-\frac{({\bf p}+{\bf k})^2}{2\mu_{\rho}}+i\epsilon } \right); \label{g+_2} \\
g_-({\bf p})  \left( \frac{1}{U_{bf}} - \frac{1}{V} \sum_{\bf k} \frac{1}{E- \frac{({\bf k}/2+{\bf p})^2}{2\mu}-\frac{k^2}{2\mu_{\rho}}+i\epsilon} \right) 
&=&\frac{1}{V} \sum_{\bf k} \left( \frac{f({\bf k})}{E- \frac{({\bf k}/2+{\bf p})^2}{2\mu}-\frac{k^2}{2\mu_{\rho}}+i\epsilon }+  \frac{g_+({\bf k})}{ E- \frac{({\bf p}-{\bf k})^2}{8\mu}-\frac{({\bf p}+{\bf k})^2}{2\mu_{\rho}}+i\epsilon } \right). \label{g-_2} 
\end{eqnarray}
By comparing the last two equations, we can see that $g_+$ and $g_-$ are identical with each other, i.e., $g_+({\bf k})=g_-({\bf k})=g({\bf k})$. This is because the interaction strength between the boson and each individual fermion is the same (both are $U_{bf}$). Finally the above three equations can be reduced to two coupled equations in terms of $f({\bf k})$(or $a_{ad}(\bf k)$) and $g({\bf k})$, as shown by Eqs.(6,7) in the main text.

\subsection{II. Validity of mean-field prediction to $a_{ad}$}

In the mean-field framework, by matching the density-density interactions between boson-fermion and between boson-dimer:
\begin{equation}
\frac{2\pi a_{bf}}{\bar{\mu}} n_b (2n_f)=\frac{2\pi a_{ad}}{\mu_{\rho}} n_b n_d ,
\end{equation}
where $n_b,\ n_f,\ n_d$ are respectively the densities of bosons, each species of fermions, dimers ($n_f=n_d$), one can get\begin{equation}
a_{ad}=\frac{4(\eta+1)}{2\eta+1} a_{bf}. \ \ \ \ \ (\eta=m_f/m_b)  \label{mf_aad}
\end{equation}
For equal mass case, $\eta=1$, we have $a_{ad}=8/3 a_{bf}$.

The exact three-body equations (Eqs.(6,7) in the main text) can reproduce above result provided the following two conditions are satisfied:

1) First, Eq.(7) in the main text should be able to produce 
\begin{equation}
g({\bf p})=\frac{2\pi a_{bf}/\bar{\mu}}{E-p^2/(2\mu)},
\end{equation}
which corresponds to approximating the terms inside the bracket of left-hand-side(lhs) of Eq.(7) as $\bar{\mu}/(2\pi a_{bf})$, and also neglecting the terms inside the bracket of right-hand-side(rhs). These approximations are valid only when $a_{bf}$ is sufficiently small, i.e., smaller than any other length scale in this problem:  $|a_{bf}|\ll 1/\kappa^*,\ a_{ff}$.

2) Secondly, $\kappa^*\gg 1/a_{ff}$ is further required in order to get Eq.(\ref{mf_aad}) from integrating the rhs of Eq.(6) in the main text. 

Combining 1) and 2), we conclude that the necessary condition for the mean-field prediction to be valid is 
\begin{equation}
|a_{bf}|\ll 1/\kappa^* \ll a_{ff}.
\end{equation}
However, since $1/\kappa^*\sim r_0$ is a very short length scale which is comparable to the range of interacting potential, above requirement means that the system cannot stay in the scale-invariant regime where all scattering lengths are much larger than $r_0$. This apparently violates (A) in deriving the generalized Efimov's radial law, and shows that the radial law (or Eq.(1) in the main text) is indeed incompatible with the mean-field prediction. For typical cold atoms systems with $r_0$ much shorter than the scattering lengths, we conclude that the mean-field prediction to $a_{ad}$ generically fails. For particular systems with $r_0$ comparable to or even larger than certain scattering length, one has to further incorporate the finite-range effect in these systems to test the validity of mean-field prediction to $a_{ad}$(Eq.(\ref{mf_aad})).

\end{widetext}


\begin{thebibliography}{99}

\bibitem{Efimov}
V. Efimov, Sov. J. Nucl. Phys. {\bf12}, 589 (1971) [{\it Yad. Fiz}. {\bf12}, 1080 (1970)]; Sov. J. Nucl. Phys. {\bf29}, 546 (1979) [{\it Yad. Fiz}. {\bf29}, 1058 (1979)].

\bibitem{Braaten}
E. Braaten and H.-W. Hammer, Phys. Rep. {\bf428}, 259 (2006).


%% need classify : ========
\bibitem{Efimov_Exp0} %Cs: loss at a<0, interference mini at a<0
T. Kraemer, M. Mark, P. Waldburger, J. G. Danzl, C. Chin, B. Engeser, A. D. Lange, K. Pilch, A. Jaakkola, H.-C. N\"{a}gerl and R. Grimm, Nature {\bf 440}, 315 (2006).

\bibitem{Efimov_Exp1}T. B. Ottenstein, T. Lompe, M. Kohnen, A. N. Wenz,
and S. Jochim, Phys. Rev. Lett. {\bf 101}, 203202 (2008). %Li6: loss peak at a<0

\bibitem{Efimov_Exp1bu}J. R. Williams, E. L. Hazlett, J. H. Huckans, R. W. Stites, Y. Zhang, and K. M. O'Hara, Phys. Rev. Lett. {\bf 103}, 130404 (2009). %Li6: loss peak at a<0

\bibitem{Efimov_Exp2}
G. Barontini, C. Weber, F. Rabatti, J. Catani, G. Thalhammer, M. Inguscio and F. Minardi, Phys. Rev. Lett. {\bf103}, 043201 (2009). %K41-Rb87: loss peak at a<0; loss peak at a>0 due to a-d resonance(without creation of molecules!)

\bibitem{Efimov_Exp3}
M. Zaccanti, B. Deissler, C. D'Errico, M. Fattori, M. Jona-Lasinio, S. M\"{u}ller, G. Roati, M. Inguscio and G. Modugno, Nat. Phys. {\bf5}, 586 (2009). %K39: loss peak at a<0; interference minima at a>0; loss peak at at a>0 due to a-d resonance(without creation of molecules!)

\bibitem{Efimov_Exp4}
N. Gross, Z. Shotan, S. Kokkelmans and L. Khaykovich, Phys. Rev. Lett. {\bf103}, 163202 (2009). %Li7: loss peak at a<0; interference minima at a>0

\bibitem{Efimov_Exp5}
S. E. Plooack, D. Dries and R. G. Hulet, Science {\bf326}, 1683 (2009). %Li7: loss peak at a<0; peak at a>0 due to a-d resonance(without creation of molecules!)

\bibitem{Efimov_Exp9}
M. Berninger, A. Zenesini, B. Huang, W. Harm, H.-C. N\"{a}gerl, F.
Ferlaino, R. Grimm, P. S. Julienne and J. M. Hutson, Phys. Rev.
Lett. {\bf107}, 120401 (2011). %Cs: loss peak at a<0 for three different FR

\bibitem{Efimov_Exp10}
R. J. Wild, P. Makotyn, J. M. Pino, E. A. Cornell and D. S. Jin,
Phys. Rev. Lett. {\bf108}, 145305 (2012). %Rb85: loss peak at a<0

\bibitem{Efimov_Exp6}S. Knoop, F. Ferlaino, M. Mark, M. Berninger, H.
Sch\"{o}bel, H.-C. N\"{a}gerl, and R. Grimm, Nat. Phys. {\bf 5}, 227
(2009).  %Cs: loss peak at a>0 due to a-d resoannce (with molecules)


\bibitem{Efimov_Exp7}
S. Nakajima, M. Horikoshi, T. Mukaiyama, P. Naidon and M. Ueda, Phys. Rev. Lett. {\bf105}, 023201 (2010). %Li6: loss peak at a>0 due to a-d resoannce (with molecules)

\bibitem{Efimov_Exp8}T. Lompe, T. B. Ottenstein, F. Serwane, K. Viering, A.
N. Wenz, G. Z\"{u}rn, and S. Jochim, Phys. Rev. Lett. {\bf 105},103201 (2010) %Li6: loss peak at a>0 due to a-d resoannce (with molecules)


\bibitem{Efimov_Exp11}R. S. Bloom, M.-G. Hu, T. D. Cumby, and D. S. Jin,
Phys. Rev. Lett. {\bf 111}, 105301 (2013). %K40-Rb87: loss peak at a>0 due to a-d resoannce (with molecules)


%O. Machtey, Z. Shotan, N. Gross, and L. Khaykovich, Phys. Rev. Lett. 108, 210406 (2012). %Li7: loss rate peak at a>0 due to a-d resonance(without creation of molecules!)

%=========

%rf Association of trimer energy:
\bibitem{rf_1}T. Lompe, T.B. Ottenstein, F. Serwane, A.N. Wenz,
G. Z\"{u}rn, S. Jochim, Science {\bf 330}, 940 (2010).
\bibitem{rf_2}S. Nakajima, M. Horikoshi, T. Mukaiyama, P. Naidon,
and M. Ueda, Phys. Rev. Lett. {\bf 106}, 143201 (2011).


%experimental confirmation of the discrete scaling symmetry
\bibitem{scaling_1}B. Huang, L. A. Sidorenkov, R. Grimm and J. M. Hutson, Phys. Rev.
Lett. {\bf 112}, 190401 (2014). %Cs
\bibitem{scaling_2}S.-K. Tung, K. Jimenez-Garcia, J. Johansen, C. V.
Parker, and C. Chin, arXiv: 1402.5943. %Li-Cs
\bibitem{scaling_3}R. Pires, J. Ulmanis, S. H¬afner, M. Repp, A. Arias, E.
D. Kuhnle, and M. Weidem¬uller, arXiv: 1403.7246. %Li-Cs


\bibitem{Salomon}I. Ferrier-Barbut, M. Delehaye, S. Laurent, A. T. Grier, M. Pierce, B. S. Rem, F. Chevy, and C. Salomon, arxiv: 1404.2548.


\bibitem{supple} See supplementary material for the detailed derivation of three-body equations and discussion on mean-field prediction of $a_{ad}$.
%According to L-S equation, the rhs of these equations are directly $\lim_{{\bf r}\rightarrow 0}$, while $1/U\sim \lim_{r\rightarrow 0} (1/r-1/a_s)$, therefore it is equivalent to BP boundary condition. An alternative way to see this is by writing $U$ in the form of pseudo-potential, then according to eq.\ref{U}, we also have such boundary condition of $\Psi$.

\bibitem{Braaten2} E. Braaten, H.-W. Hammer, D. Kang, and L. Platter, Phys. Rev. A {\bf 81}, 013605 (2010).

\bibitem{footnote_loss}For the case $x\in(0,1)$, as in the ENS experiment which was mostly carried out in the $a_{bf}<|a_{ff}|$ regime\cite{Salomon}, $g({\bf k})$ (Eq.(\ref{eq2})) will have a pole at finite ${\bf k}$ and both $f({\bf k})$ and $g({\bf k})$ will become complex. We thus expect the Bose and Fermi mixtures in ENS setup will suffer from three-body loss due to the decay of fermion-fermion dimer to boson-fermion dimer through three-body collisions. 


\bibitem{Eint} Here the formulation of ${\cal E}$
%are formulated by the density-density interaction, which 
requires all three scattering lengths be sufficiently small. It cannot describe the situation when $a_{ad}$ is near resonance, but can give a qualitative picture about how the change of $a_{ad}$ (according to Eq.(1)) affects the stability of the system. 


\bibitem{Petrov} D. S. Petrov, C. Salomon and G.V. Shlyapnikov, Phys. Rev. Lett. {\bf 93}, 090404 (2004).

\bibitem{book} C. J. Pethick and H. Smith, {\it Bose-Einstein Condensation in Dilute Gases} (Cambridge University Press, 2002). 

\bibitem{footnote1} As the first attempt to explore the many-body effect of Eq.(1), here we are only interested in the stability of a homogeneous mixture against density fluctuations. The phase diagram can be further improved by considering the possibility of a separation/collapse of partially polarized states. There states are possible to exist near the MIX-PS/CL boundaries, which are not addressed here. 

\bibitem{footnote} Near the PS-CL phase boundary, $a_{ad}$ goes across a resonance, which is associated with a trimer state emerging from the atom-dimer continuum and the system will suffer an enhanced atom loss near this regime\cite{Efimov_Exp6,Efimov_Exp7,Efimov_Exp8,Efimov_Exp11}.

%\bibitem{footnote2} When $a_{ff}<a_{bf} \ (x>1)$, the fermion-fermion dimer becomes the ground state dimer and the atom loss can only be enhanced at the atom-dimer resonance (see Fig.4(a)). 

\bibitem{Chin} C. Chin, R. Grimm, P. Julienne and E. Tiesinga, Rev. Mod. Phys. {\bf 82}, 1225 (2010).



\end{thebibliography}
\end{document}